# Fundamental physical and resource requirements for a Martian magnetic shield

Marcus DuPont[1,2]  and Jeremiah W. Murphy[2]

[1]Department of Physics, New York University, New York, NY 10001, USA and [2]Department of Physics, Florida State University, Tallahassee, FL 32306, USA

## Abstract

Mars lacks a substantial magnetic field; as a result, the solar wind ablates the Martian atmosphere, and cosmic rays from solar flares make the surface uninhabitable. Therefore, any terraforming attempt will require an artificial Martian magnetic shield. The fundamental challenge of building an artificial magnetosphere is to condense planetary-scale currents and magnetic fields down to the smallest mass possible. Superconducting electromagnets offer a way to do this. However, the underlying physics of superconductors and electromagnets limits this concentration. Based upon these fundamental limitations, we show that the amount of superconducting material is proportional to $B_c^{-2} a^{-3}$, where $B_c$ is the critical magnetic field for the superconductor and $a$ is the loop radius of a solenoid. Since $B_c$ is set by fundamental physics, the only truly adjustable parameter for the design is the loop radius; a larger loop radius minimizes the amount of superconducting material required. This non-intuitive result means that the 'intuitive' strategy of building a compact electromagnet and placing it between Mars and the Sun at the first Lagrange point is unfeasible. Considering reasonable limits on $B_c$, the smallest possible loop radius is ∼10 km, and the magnetic shield would have a mass of $\sim 10^{19}$ g. Most high-temperature superconductors are constructed of rare elements; given solar system abundances, building a superconductor with $\sim 10^{19}$ g would require mining a solar system body with several times $10^{25}$ g; this is approximately 10% of Mars. We find that the most feasible design is to encircle Mars with a superconducting wire with a loop radius of ∼3400 km. The resulting wire diameter can be as small as ∼5 cm. With this design, the magnetic shield would have a mass of $\sim 10^{12}$ g and would require mining $\sim 10^{18}$ g, or only 0.1% of Olympus Mons.

## Introduction

Long-term terraforming of Mars would require the revival of its atmosphere, which over long timescales also requires protective properties of a global magnetic field. In the past, Mars most likely had an atmosphere and an ocean comparable to Earth's Arctic Ocean (Villanueva et al. 2015). It also had a magnetosphere to protect the atmosphere (Connerney et al. 1999), and the atmosphere protected the ocean. In addition to protecting the atmosphere, a magnetosphere stops cosmic rays from reaching the planet's surface. Unimpeded, these cosmic rays would damage DNA and other bio-organic molecules (e.g. Baumstark-Khan and Facius 2002; Dartnell et al. 2007). Therefore, an important step in terraforming Mars is building an artificial magnetosphere, a magnetic shield.

Understanding the evolution of Mars' atmosphere and ocean helps to motivate the need for a magnetic shield. Currently, the atmospheric pressure at the base of Mars is 0.01 bar, roughly 1% of Earth atmospheric pressure. In addition, the Martian atmosphere has very little oxygen (0.06%). Even if the Martian atmosphere had Earth-like oxygen percentages (21%), the atmospheric pressure is just below what is survivable by humans (0.063 bar) – this is known as the Armstrong limit (NAHF 2007). If the desire is to breathe a Martian atmosphere, then the Martian atmosphere needs to be thicker.

The thickness of the Martian atmosphere also has consequences for the long-term viability of a Martian Ocean. Geology and geochemistry measurements (Villanueva et al. 2015) reveal the current D/H ratio in water on Mars to be on the order of $10^{-5}$, which is higher than previously postulated primordial ratios (Usui et al. 2012). The high concentration of D/H today suggests that Mars may have had an ocean as large as the Arctic Ocean (Villanueva et al. 2015). Atmospheric models suggest that such an ocean can persist with a substantial Martian atmosphere (Read and Lewis 2004; Fairén et al. 2009; Fairén 2010; Fairén et al. 2011). Yet, today the Martian atmosphere is so thin that any engineered ocean would sublime and then evaporate into space on a time scale of $\sim 10^4$ years (Carr and Head 2003).

Current theories suggest that the collapse of the Martian atmosphere was likely precipitated by the collapse of the Martian dynamo and magnetic field (Acuña et al. 1999). Even though the Martian magnetic field is quite weak (0.015 G), the Mars Global Surveyor (MGS) measured magnetized crust in excess of 0.6 G (Acuna et al. 1998). This is comparable to Earth's surface magnetic field of 0.5 G, suggesting that the Martian crust was magnetized by a historic Martian dynamo. It is likely that this ancient Martian magnetosphere deflected the solar wind and protected the Martian atmosphere (e.g. Johnson et al. 2020). However, Mars lost that natural magnetosphere, leaving the Martian atmosphere vulnerable to ablation by the solar wind. Measurements made by the *Mars Atmosphere and Volatile Evolution* (MAVEN) spacecraft indicate significant past and current outgassing. One indication is that 66% of argon on Mars has been lost to space (Jakosky et al. 2018). In addition, the MAVEN team estimates that the outgassing rate for H and O is 2–3 kg s$^{-1}$. According to Jakosky et al. (2018), this current mass loss rate would only remove about ∼ 78 mbar of $CO_2$ and a few metres global equivalent of water from the Martian environment. This current rate is not enough to remove a Martian atmosphere with 1 bar of pressure. However, the solar wind was significantly stronger in the past due to higher rotation rates and solar activity, and Jakosky et al. (2018) estimate that ∼ 0.8 bar of $CO_2$ and the equivalent of a 23 m global ocean have ablated away over the last 4 Gyr. This catastrophic atmospheric loss is likely due to the collapse of the Martian dynamo 4 Gyr ago. Therefore, to protect any Martian atmosphere building attempt, humankind will need to build an artificial magnetosphere to protect the nascent atmosphere from the solar wind.

Protecting human habitation with magnetic shields is not a new idea. For example, on a much smaller scale, Bamford et al. (2014) designed spacecraft-scale magnetospheres to protect human occupants from energetic solar particles during deep space travel. Although the planetary-scale design in this paper may have implications for efficient spacecraft-scale designs, we do not focus on those here. See the 'Discussion' section for those implications.

To protect Mars from the solar wind, Green et al. (2017) considered placing a magnetic shield at the L1 Lagrange point in the Sun–Mars system. Such a spacecraft would always be in between the Sun and Mars and given an adequate magnetic field would protect Mars from the solar wind. They modelled the interaction of the solar wind with a magnetic shield to validate that an artificial magnetosphere could, in principle, be used to effectively protect Mars. They found that an electromagnet with large scale magnetic fields comparable to Earth's is sufficient to shield Mars.

Even if construction of a magnetic shield is possible, there are challenges in building up an atmosphere. Mars is currently outgassing to space, but stopping this outgassing is not enough to build a Martian atmosphere on human timescales. Barring any losses to space, it would take $10^7$ years to double the mass of the atmosphere (Jakosky and Edwards 2018). Additionally, the available $CO_2$ at Mars is insufficient due to it only being able to produce a maximum atmospheric pressure of 20 mbar if released (Jakosky and Edwards 2018), making natural outgassing impractical. Mars requires both sublimation of polar deposits and imported outside volatiles that could scale the outgassing rate appropriately if significant greenhouse warming is to take effect. Building a Martian atmosphere is the subject for another paper.

In this paper, we calculate the fundamental physical and resource requirements for building a Martian magnetic shield. The fundamental challenge is to compress planetary scale magnetic fields and currents into as small as a volume as possible. As we will show, the physics and resource constraints challenge the physics of superconductors and the abundance of materials in the Solar System that could be used to build a superconducting magnetic shield. Our primary focus is to derive the important scales and scaling relationships based upon the physics of superconductors, the physics of electromagnets and the abundance of materials in the Solar System.

### Deriving constraint equations for a magnetic shield

The strategy for deriving the constraint equations are as follows. In the first step, we estimate the magnetic field required to deflect the solar wind. Since Earth's magnetosphere represents a successful example of deflecting the solar wind, we use scaling relations to estimate the magnetic field required to deflect the solar wind at Mars' orbit. In the second step, we use this required magnetic field and the critical magnetic field of superconductors to derive the minimum mass of the superconducting electromagnet. The third step is to estimate the abundance of the rarest element of the superconducting material. Finally, we use these constraints to propose a design for the magnetic shield that maximizes its ability to deflect the solar wind with the least amount of rare superconducting material.

One possible strategy is to build the most compact superconductor, using the fewest resources. Due to fundamental physics of superconductors, there are limits to the compactness. Superconductors rely on Cooper pairs to transport current, and these Cooper pairs have a finite binding energy. These Cooper pairs remain bound below a critical temperature and/or below a critical magnetic field. If the magnetic field is too high in the superconductor, then the magnetic field can disrupt the Cooper pairs – pairs of electrons (or fermions) bound at low temperatures. Once these Cooper pairs are disrupted (through heating or applied magnetic fields), the electrons can no longer occupy the same quantum state, quenching the superconductor. The sudden increase in resistance to the current causes significant heating and can lead to explosive consequences. Hence, superconductors are limited to a critical magnetic field and temperature, $B_c$ and $T_c$.

Therefore, the two constraints on the electromagnet are that (1) the magnetic field must be strong enough to deflect the solar wind around Mars, but (2) it must be weaker than the critical magnetic field, $B_c$. Using these constraints and seeking to minimize the solenoid mass, we determine the wire bundle radius, $d$, and the loop radius, $a$ of the electromagnet. After deriving the ideal geometry of the device, we calculate the power required to keep the superconductor below its critical temperature, $T_c$. Given the minimum mass of the superconductor and solar system abundances, we then estimate the mass of bulk rock required to extract enough raw material for the magnetic shield.

### Minimum volume of the magnetic shield

The first step in estimating the minimum volume is to estimate the minimum magnetic field necessary to deflect the solar wind at Mars' orbit. To arrive at the minimum field required, we relate the ram pressure of the solar wind to the magnetic pressure at Mars. The magnetic field will deflect the solar wind when the magnetic pressure, $B(r)^2/8\pi$ is of order the ram pressure of the solar wind, $P_{\text{ram}} \sim \rho v^2$:

$$\rho v^2 \approx \frac{B(r)^2}{8\pi}. \tag{1}$$

In these expressions, ρ is the density of the solar wind, $v$ is the wind speed and $B$ is the magnitude of the magnetic field. Earth's magnetopause successfully deflects the solar wind; therefore, we derive the requirements on the strength of the Martian magnetic shield by scaling the conditions of Earth's magnetopause. A key parameter in the upcoming calculations is $r_s$, the standoff radius. The standoff radius is the distance from planet centre to the magnetopause, so a careful choice of this parameter affects the field intensity required. If we assume a dipole field for Earth's magnetic field, then the balance in pressures, equation (1), is

$$P_{\text{ram},\oplus} \approx \frac{B(r)^2}{8\pi} = \frac{B_\oplus^2}{8\pi}\left(\frac{R_\oplus}{r_s}\right)^6 \quad (2)$$

where $R_\oplus$ is the Earth radius, $r_s$ is the standoff radius of Earth's magnetopause and $B_\oplus$ is Earth's average magnetic field strength at the surface. The ram pressure of the solar wind as a function of distance, $D$, is $P_{\text{ram}} \approx \rho v^2$. The mass loss rate from the Sun is $\dot{M} = \rho 4\pi D^2 v$. Assuming that $\dot{M}$ and $v$ are constant at the distances of Earth and Mars, the ram pressure as a function of distance is

$$P_{\text{ram}} \approx \frac{\dot{M}v}{4\pi D^2} \Rightarrow P_{\text{ram},\mars} = P_{\text{ram},\oplus}\left(\frac{D_\oplus}{D_\mars}\right)^2,$$

where $D_\oplus$ is the distance from the Sun to Earth, and $D_\mars$ is the distance from the Sun to Mars. Now, having related the ram pressure to both the magnetic field and the distance from the Sun, we calculate the required strength of the Martian magnetic shield. Once again, we assume a dipole magnetic field and scale the strength of the magnetic field at the surface of Mars: $B_{\text{shield}} = B_0(R_\mars/r'_s)^3$, where $R_\mars$ is the radius of Mars and $r'_s$ is the standoff radius of the Martian magnetopause. This yields,

$$B_0^2\left(\frac{R_\mars}{r'_s}\right)^6 = B_\oplus^2\left(\frac{R_\oplus}{r_s}\right)^6\left(\frac{D_\oplus}{D_\mars}\right)^2. \quad (3)$$

Roughly, $R_\mars \approx R_\oplus/2$ and $D_\mars \approx (3/2)D_\oplus$. Given these, the strength of the Martian magnetic shield is

$$B_0 = \frac{16}{3}B_\oplus\left(\frac{r'_s}{r_s}\right)^3. \quad (4)$$

Once again, $r'_s$ is the standoff radius of Mars' magnetosphere, a free parameter, and $r_s$ is the standoff radius for Earth's magnetopause, $r_s = 6R_\oplus$. Note that the minimum field strength depends on a choice of the Martian magnetopause, $r'_s$. To calculate the minimum requirements for the magnetic field, we choose a relatively small standoff radius for the Martian magnetopause. The absolute minimum would be $r'_s = R_\mars$. However, such a small standoff radius would barely protect the Martian atmosphere, so we choose a slightly larger radius. To provide a little buffer, we choose $r'_s = 2R_\mars$, which is $r'_s \approx R_\oplus$, since $2R_\mars \approx R_\oplus$. This choice also simplifies the ratio of standoff radii, $r'_s/r_s \approx 1/6$. Given these choices, the minimum field strength is $B_0 \approx B_\oplus/40$, which corresponds to a magnetic field strength of 1.25 μT at the Martian surface. This simple order-of-magnitude estimate for the necessary magnetic field strength is consistent with the results of magnetohydro simulations (Green et al. 2017). Next, we calculate the mass of the magnetic shield given $B_0$ and $B_c$. In general, the magnetic field is a function of distance, $B(r)$. At the edge of the superconducting magnet, the magnetic field can be no larger than $B_c$, yet the magnetic must be have a characteristic strength larger than $B_0$ to deflect the solar wind. The design of the Martian magnetic shield is fundamentally a solenoid with $N$ loops, each with a loop radius of $a$; the bundle of wires has a radius $d$. In the following derivation, we show that the constraints on the magnetic field impose constraints on the dimensions for the solenoid. For an illustration of the solenoid and magnetic field strength, see Fig. 1. Since the goal is to compress planetary scale currents and magnetic fields into as small a volume as possible, this solenoid will need to be made with superconducting wires. Using Biot–Savart's law, the magnetic field in the plane of the solenoid a distance $r$ from the centre is

$$B(r) = -\frac{\mu_0 N I_c}{4\pi a}\int_0^{2\pi}\frac{((r/a)\cos\theta - 1)d\theta}{(1 + (r/a)^2 - (2r/a)\cos\theta)^{3/2}} \\
= -\frac{\mu_0 N I_c}{4\pi a}f\left(\frac{r}{a}\right). \quad (5)$$

The prefactor sets the overall scale of the magnetic field, and the dimensionless integral gives the radial dependence. In general, the integral, $f(r/a)$ does not have a closed form, so we explore limiting cases to show that this equation reduces to well-known fields. These limiting cases are also useful in providing an analytic constraint equation for the magnetic shield. There are three limiting cases that are analytic and are useful in the design of the magnetic shield. First, at the centre of solenoid, $r = 0$, the magnetic field in equation 5) is

$$|B(0)| = \frac{\mu_0 N I_c}{2a}. \quad (6)$$

Second, in the far-field limit, $r \gg a$, $B(r)$ approaches the dipole approximation

$$\lim_{r \gg a}|B(r)| \approx B(0)\left(\frac{a}{r}\right)^3. \quad (7)$$

Third, for a solenoid with a large loop radius $a$ (bundle radius $d$ is much smaller than $a$) the magnetic field just outside the wire bundle approaches the limit for an long straight wire,

$$\lim_{d \ll a}|B(a+d)| \approx \frac{\mu_0 N I_c}{2\pi d}. \quad (8)$$

Equations (6)–(8) represent three useful limiting cases that we will use later in the derivation. Before we use the approximate magnetic field expressions, we first use the exact expression, equation (5), to find algebraic expressions for the magnetic field constraints. For the first condition, the magnetic field at the surface of the wire bundle should be less than the critical magnetic field, $B(a+d) < B_c$. In this case, the exact equation, equation (5), becomes

$$|B(a+d)| = \frac{\mu_0 N I_c}{4\pi a}f\left(1 + \frac{d}{a}\right) = B_c. \quad (9)$$

For the second condition, the magnetic field at a standoff radius of $r_s = 2R_\mars$ is the minimum ($B_0$) required to deflect the solar

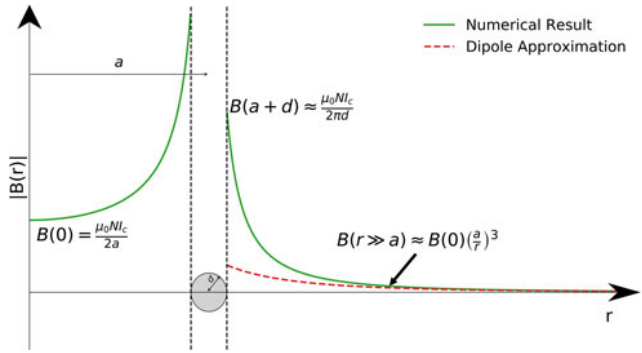

**Fig. 1.** Magnetic field strength as a function of distance from the centre of the solenoid. The solid curve represents the numerical integration of equation (5). We also show three standard approximations for solenoids. The central magnetic field, $B(0)$, is roughly equal to the core magnetic field for a current loop. The magnetic field at the surface of the wire is roughly that of a straight wire. The dashed line is the far-field dipole approximation.

wind, $B(2R_{\male}) = B_0$. In this case, the exact equation, equation (5), becomes

$$|B(2R_{\male})| = \frac{\mu_0 N I_c}{4\pi a} f\left(\frac{2R_{\male}}{a}\right) = B_0. \quad (10)$$

The approximate analytic equivalents of the these two constraint equations are as follows. The equivalent of equation (9) is

$$|B(a+d)| \approx \frac{\mu_0 N I_c}{2\pi d} \approx B_c, \quad (11)$$

and the equivalent of equation (10) is

$$|B(2R_{\male})| \approx \frac{\mu_0 N I_c}{2a}\left(\frac{a}{2R_{\male}}\right)^3 \approx B_0. \quad (12)$$

Together, the exact constraint equations, equations (9) and (10), provide a joint constraint on the loop radius, $a$, and the wire bundle radius, $d$. Solving both equations for $\mu_0 N I_c/(4\pi a)$ and then equating them gives the following constraint

$$\frac{f(2R_{\male}/a)}{f(1+d/a)} = \frac{B_0}{B_c}. \quad (13)$$

In summary, this constraint defines the magnetic shield dimensions required to achieve the magnetic field ratio, $B_0/B_c$. It is a two parameter constraint that depends upon $a$ and $d$. For a given loop radius, $a$, one may find the wire bundle radius, $d$, that satisfies the constraint. Again, due to equation (5) not having a closed form for intermediate values of $r$, we used numerical methods to both compute the magnetic field (Fig. 1) and the wire bundle radius, $d$ (Fig. 2). Using the approximations for the magnetic field near the wire (equation (11)) and far from the wire equation (12), we derive an analytic expression for equation (13):

$$\frac{B_0}{B_c} = \frac{\pi d a^2}{8 R_{\male}^3} \quad (14)$$

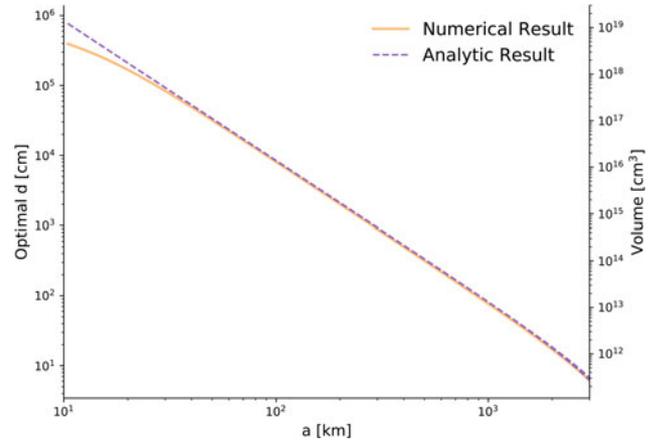

**Fig. 2.** Physical constraints on the solenoid dimensions: wire radius ($d$) and superconducting material volume as a function of loop radius ($a$). The dashed purple line shows the analytic result, equation (7), and the solid yellow curve is the numerical result, equation (13). These dimensions are calculated by requiring the field strength to be $B(2R_{\male}) = B_{\oplus}/40$ and below a critical superconducting field of 200 T at the wire. The left axis shows the superconducting bundle radius ($d$) as a function of the solenoid radius ($a$); the right axis shows the material volume ($2\pi^2 d^2 a$) as a function of $a$. The required volume goes down for a wider loop. Therefore, the most optimum solution is to wrap the magnetic shield around Mars.

The material volume is dependent on the dimensions of the loop bundle as well as the loop radius,

$$V_{\text{superconductor}} = 2\pi^2 d^2 a \quad (15)$$

where we calculate $d$ using the constraint equation, equation (13). Using the analytic constraint, equation (14), we can express the volume in terms of the magnetic field constraints and the loop radius,

$$V_{\text{superconductor}} = 128\left(\frac{B_0}{B_c}\right)^2 \frac{R_{\male}^6}{a^3}. \quad (16)$$

This equation shows that the physics of superconductors and solenoids apply strong constraints on the design of the magneto-shield. For one, the superconducting volume goes down as the critical magnetic field squared. The strongest dependence is on the loop radius. The material required for solenoid goes down as the loop radius cubed. Figure 2 shows both $d$ and $V$ as a function of $a$. The solid line shows the numerical result of solving equation (13) for $d$; the dashed line shows the result when using the analytic result, equation (14). As a reminder, constraints on the magnetic field are that $B_0 = B_{\oplus}/40$, and $B_c = 200$ T. The critical field value is of order the critical magnetic field for bismuth strontium calcium copper oxide (BSCCO), which was discovered to have one of the largest $B_c$ for a superconducting wire (Golovashkin et al. 1991). Another unique quality about BSCCO is that it is able to produce round wires, ideal for this type of shielding construction. Based on this result, if we wrapped a wire around Mars' equator (operating near the critical field of ~200 T), the material volume would be roughly $2.25 \times 10^5$ m$^3$. Moreover, a smaller loop radius $a$ requires a much larger wire thickness and hence volume of superconductor. For example, a loop radius of $a \sim 10$ km, would require a wire radius of $d \sim 10$ km, giving a volume of $\sim 10^{12}$ m$^3$. An illustration of the different geometries we consider for the magnetic shield is shown in Fig. 3.

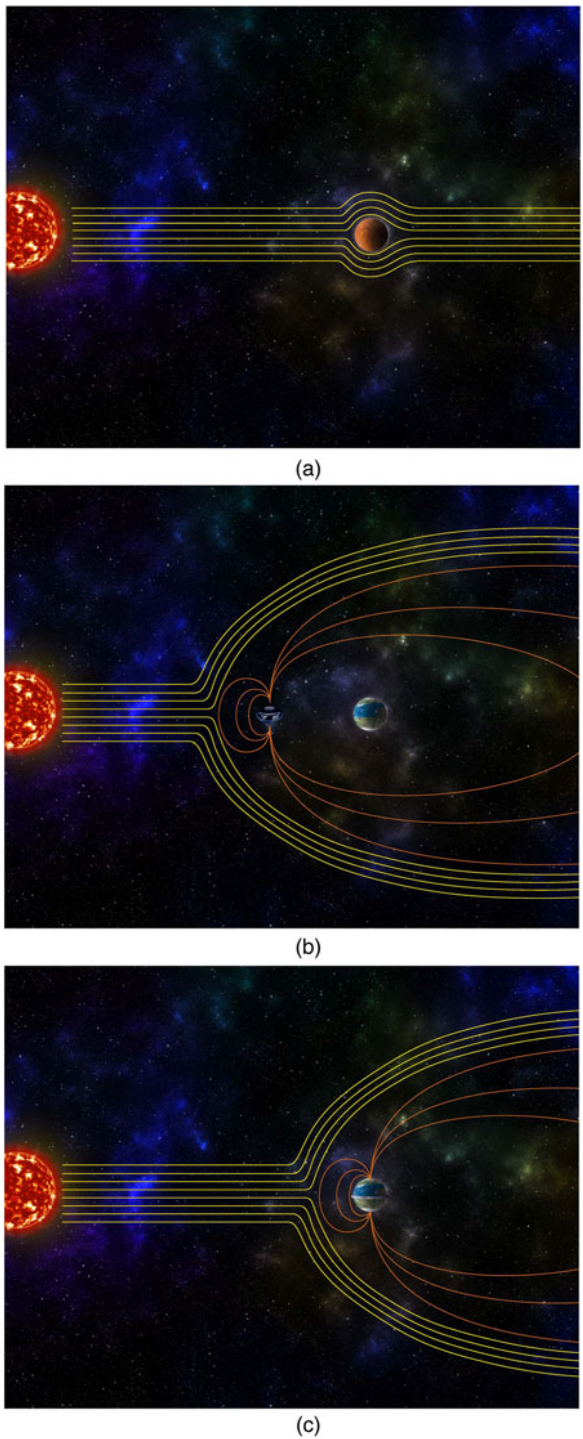

(a)

(b)

(c)

**Fig. 3.** Martian magnetic shield designs. Panel (a) shows the current interaction between the Martian atmosphere and the solar wind. A lack of magnetic field allows the solar wind to ablate away the Martian atmosphere, making Mars inhospitable to life. The challenge of building a Martian magnetic shield is to build an electromagnet that can produce planetary scale magnetic fields with the least amount of superconducting material; however, the critical magnetic field of superconductors limits the size of the electromagnet. The mass of the electromagnet scales as $B_c^{-2} a^{-3}$, where $B_c$ is the superconducting critical magnetic field, and $a$ is the loop radius. Panel (b) shows the most compact design possible: a solenoid with a wire radius and loop radius of order 10 km. Counter-intuitively, this 'compact' design would require the most superconducting material and would require mining 10% of Mars to extract the rare superconducting material. This is clearly an impractical design. Panel (c) shows a more practical design: the electromagnet is a loop of wire around the Martian equator. The wire radius is $b = 5$ cm and requires mining only 0.1% of Olympus Mons for the rare superconducting material.

### Power requirements to keep the conducting wire cool

We now visit the power needed to maintain the superconductor at a temperature of order 100 K. The optimal electromagnet design is a superconducting wire around Mars that has a wire diameter of 5 cm. Insulating the superconducting wire will require building an insulating tube around the wire. No insulator is perfect and some heat will leak into the system. As a result, there will need to be active cooling systems placed along the wire, and these cooling systems will require some power to maintain the low temperatures. Typically the power required is of order the power of heat flowing through the insulator. For the purposes of this order-of-magnitude calculation, we imagine that the insulator is a 1 m diameter tube that envelopes the entire loop. The tube has a near vacuum. Except for the loop radius being the size of Mars, the other dimensions (tube diameter) and design are similar to standard dewars (vacuum flasks). The power of heat flow through an insulator is given by the Fourier equation for unidirectional conduction

$$\dot{Q} = \frac{A}{L} \int_{T_0}^{T} k(T') dT' = \frac{A}{L} \bar{k} \Delta T \qquad (17)$$

where $A$ is the cross-sectional area, $L$ is the length of the heat conduction path, $\dot{Q}$ is the heat flow rate, $T$ is temperature, $k$ is the thermal conductivity of the material and $\bar{k} = \int_{T_0}^{T} k(T') dT'/\Delta T$ is the mean thermal conductivity of the material. Vacuum-insulated panels can achieve thermal conductivity as low as 0.004 W m$^{-1}$ K$^{-1}$ (Véjelis et al. 2010). Using this value, a temperature difference $\Delta T = (T_{ambient} - T_c) \sim 173$ K, a cross-sectional area of $10^7$ m$^2$, and a conduction path of 1 m, we obtain a vacuum dewar heat flow rate of 1 MW. This is the average power capacity of a wind turbine. Hence, the primary challenge in constructing the electromagnet is not in cooling it.

### Can the device fit in a rocket? Critical magnetic field required to fit the electromagnet on today's rockets

To further illustrate the extreme constraints for building the Martian magneto-shield, we calculate what the required $B_c$ would be if one built the superconducting wire on Earth, shipped it in a rocket and placed it a the L1 Lagrange point. This strategy requires fitting the wire inside the rocket fairing. Once completed, the SpaceX Starship will have the largest payload capacity. Starship has a 9 m diameter; for ease of calculation we will just assume that today's current rockets could fit a coil of superconductor that has a volume of $10^3$ m$^3$. If one kept the superconductor compact with $a \sim d \sim 10$ m, then equation (14) indicates that the critical magnetic field would have to be $10^{11}$ T or $10^{15}$ G. This is a magnetar field, and the energy density of such a field far exceeds the bulk modulus of any terrestrial solid. Clearly, this solution is not feasible. Another alternative is to transport the spool to Mars, unwind it to make a loop around Mars. With this configuration, the wire radius would have to be $d \sim 7$ mm, and the critical magnetic field would have to be $B_c \sim$ 1400 T. This is a factor of ten larger than current $B_c$ limits. Fitting the superconducting material in one of today's rockets is not feasible either. These estimates show that it is not possible to build the electromagnet on Earth and transport it to Mars. Rather, the electromagnet must be constructed on or near Mars using local material.

### *Bulk rock mass required to mine superconducting material*

Using the solar abundances shown in Fig. 4 (Cameron 1973), we estimate the amount of bulk rock required to mine the elemental ingredients of the superconductor. For BSCCO superconductors, the rarest element is bismuth (Bi), so its abundance would dictate the bulk rock required. Asteroids are often dominated by carbonaceous materials, so the relative ratio of rock to Bi is roughly given by the abundance ratio of C to Bi. The larger planets are predominantly silicates, of which silicon is the most abundant element. Therefore, the ratio of Si to Bi gives a rough scaling for the ratio of rock to Bi. For the optimum magneto-shield design (loop around Mars' equator), the superconductor volume is $2 \times 10^5$ m$^3$. Assuming an average rock density of 3 g cm$^{-3}$, the superconducting mass is $6 \times 10^{11}$ g. The ratio of mass in Bi to the bulk rock mass is given by the equivalent abundance ratios:

$$\frac{M_{Bi}}{M_{C/Si}} = \frac{X_{Bi}}{X_{C/Si}} \quad (18)$$

Here, $M_{Bi}$ and $M_{C/Si}$ are the Bi ore masses and silicate/carbon masses, respectively. Given these ratios, the mass of bulk rock as a function of the loop and bundle radii is:

$$M(a, d) = 2\pi^2 \rho a d^2 \frac{X_{C/Si}}{X_{Bi}}. \quad (19)$$

Figure 5 shows the required superconductor mass and bulk rock mass as a function of the loop radius, $a$. In addition, to Bi, we also calculate the bulk rock required if the limiting element for the superconductor is yttrium or carbon. Yttrium is the limiting element for yttrium barium copper oxide (YBCO), which is a ceramic superconductor that has a high critical temperature (~93 K) and magnetic field (Wu et al. 1987) but is usually constructed in an anisotropic tape form as opposed to a round wire (e.g. Kametani et al. 2015). Carbon is included for the case of carbon nanotubes; the benefit to carbon nanotubes is that carbon is many orders of magnitude more abundant than most superconducting materials, but the drawback is that carbon nanotubes are difficult to construct long wires and they have a low critical temperature (~5 K) and magnetic field (~4 T). Figure 5 clearly indicates that the most feasible magneto-shield is one with a loop radius of order the size of the planet. If the loop radius is of order ~10 km and one fabricates the superconductor using Bi, then the amount of mined bulk rock is 10% of Mars. Even if the magneto-shield is composed of carbon nanotubes and a more common element (C), the amount of bulk rock is equivalent to the mass of a very large ~10 km asteroid. On the other hand, if the loop radius is $a \sim 10^3$ km, then the amount of bulk rock required for Bi is 0.1% of Olympus Mons[1], the largest mountain on Mars. In the case of carbon nanotubes, the bulk rock is only $10^{-6}$ of Olympus Mons. Therefore, building a loop around Mars is the most feasible solution when constructing a Martian magnetosphere.

---

[1]Note that we use Olympus Mons only to illustrate the scale. Mars' surface materials are not carbonaceous, but closer to volcanic basalt, so in practice one cannot use the materials in Olympus Mons to build this magnetic shield as they will be even further from carbonaceous. See more in Blake et al. (2013).

### Discussion and conclusion

The primary results and conclusions are as follows. First, to protect the Martian atmosphere, the magnetic shield should have a field strength of 1.25 μT at a distance of one Earth radius. This result is consistent with magneto-hydrodynamic simulations (Green et al. 2017), where we take into account the decrease in ram pressure at Mars. The binding energy of Cooper pairs limit the magnetic field strength within the superconductor to a critical magnetic field, $B_c$. Given this constraint and the physics of solenoids, we derive the total amount of material required to construct the magnetic shield (see Fig. 5). The total amount of superconducting material scales as $d^2 a$, where $d$ is the solenoid's wire bundle radius, and $a$ is the loop radius of the solenoid. Limiting the magnetic field at the surface of the solenoid to $B_c$ limits the wire bundle radius; this limit scales as $d \propto B_c^{-1} a^{-2}$. Therefore, the total amount of superconducting material scales as $B_c^{-2} a^{-3}$. In other words, the amount of superconducting volume goes down as the cube of the loop radius. Given these physical constraints, making the solenoid as compact as possible – where $d \sim a$ – is unreasonable. For a BSCCO superconductor with a $B_c$ of 200 T, the minimum size possible for a magnetic shield is $a \sim d \sim 10$ km. This corresponds to $10^{19}$ g of superconducting material. The rarest element in BSCCO superconductors is Bi, which has an abundance ratio of 0.143 when normalized to Si = $10^6$. To extract the necessary Bi, a terraforming operation would have to process $10^{25}$ g or 10% of Mars' mass. Clearly, building the most compact magnetic shield is impractical. However, a magnetic shield with a loop radius, $a$, that is the size of Mars is feasible. Since the superconducting volume goes down as the cube of $a$, the total amount of Bi required is only $\sim 5 \times 10^{18}$ g. To extract this amount of Bi would require mining only $\sim 10^{-3}$ of Olympus Mons. If the superconductor is made of carbon nanotubes, then the main constituent, carbon, is much more abundant. Constructing a magnetic shield of carbon nanotubes with a loop radius the size of Mars would only require mining $\sim 10^{-10}$ of Olympus Mons. Rather than placing the magnetic shield at the L1 Lagrange point of the Sun–Mars system, we recommend wrapping a superconducting wire around Mars. Although constructing a magnetic shield with a loop radius the size of Mars is far more feasible than $a \sim 10$ km, constructing such a long superconducting wire would have its own challenges. For instance, BSCOO is currently made in kilometre lengths. Assuming one could build a BSCOO wire 20,000 km in length operating at 200 T with $d \sim 7$ mm, the sheer processing and production costs at a target high-temperature superconductor price of \$10/kA $m$ could exceed $\sim \$\,70,000$ m$^{-1}$ using our shield estimates (Grant and Sheahen 2002). Note that this does not include the accompaniment of the vacuum insulator which will have to be wrapped around the Martian equator, near Olympus Mons as well as the caldera deep in its summit. Nonetheless, measurements indicate that the resistance in a superconducting wire is truly zero, i.e. it takes zero voltage to maintain the current in a superconducting wire. Hypothetically, a superconducting current could persist indefinitely. Experiments have confirmed that a superconducting current can persist for at least 100,000 years (File and Mills 1963). Whether this could be maintained over thousands of kilometres is completely untested. If superconducting wire is lossless over thousands of kilometres, then the power losses of a superconductor are minimal. The main source of power may be in keeping the superconducting wire at superconducting temperatures. Shi et al. (2012) investigated the superconducting properties of double-walled carbon nanotubes and yielded promising results, but one of the hurdles to overcome is figuring out how to maintain their superconducting states for extended periods. Once the magnetic shield is in place, then the next phase of the terraforming operation may begin: restoring the Martian atmosphere. The Mars Atmosphere and Volatile Evolution (MAVEN) spacecraft measured Martian outgassing of hydrogen and oxygen at a rate of $\sim 2-3$ kg s$^{-1}$ (Jakosky et al. 2018). This is equivalent (by weight) to 2 l of water leaving every second. With a magnetic shield this outgassing would begin replenishing the Martian atmosphere. At this rate, it would take a Gyr to replenish the Martian atmosphere to one equivalent to Earth's. Therefore, replenishing the Martian atmosphere will also require enhanced outgassing efforts. In summary, terraforming Mars will require building a magnetic shield to protect a nascent Martian atmosphere. Given the physics of superconductors, we show that volume of superconducting material scales as $B_c^{-2} a^{-3}$. Considering the abundance of superconducting materials, the most feasible solution is to wrap Mars in a superconducting wire.

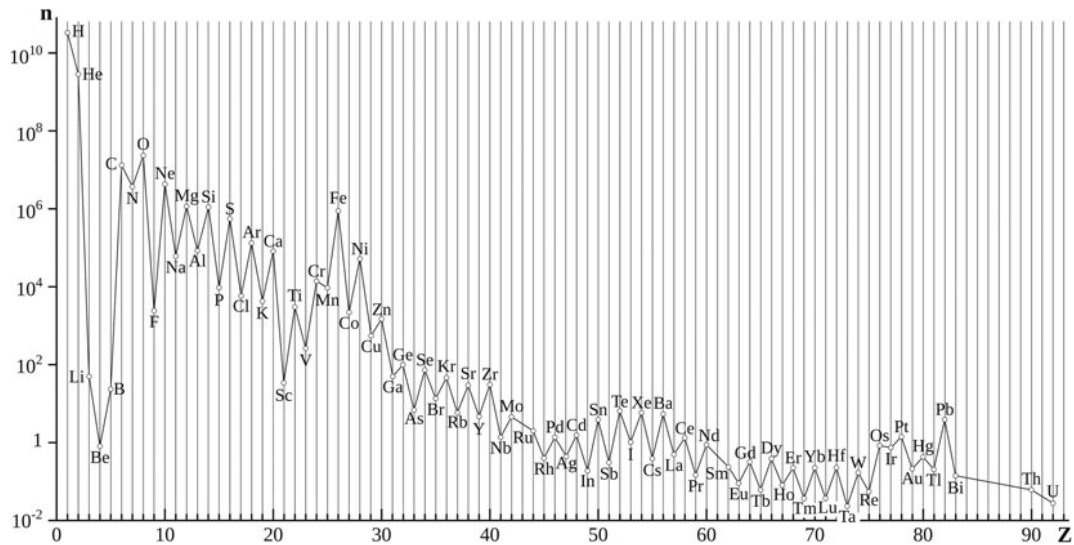

**Fig. 4.** Relative abundance of chemical elements in the Solar System (Cameron 1973). The *x*-axis represents atomic number and the *y*-axis represents abundance of elements for every $10^6$ atoms of Si. The relevant elements for this paper are Si, a primary constituent in silicate bodies such as Earth, and C, a primary constituent of most asteroids. The rare elements Bi and Y represent two important elements in the construction of high $T_c$ superconductors (e.g. the magnetic shield). Figure credit: public domain.

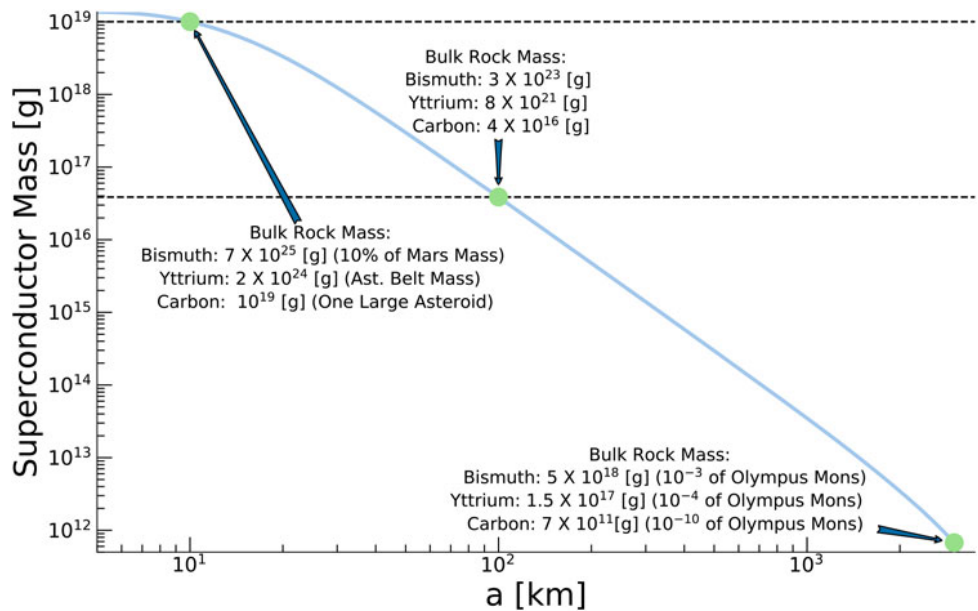

**Fig. 5.** Constraints on the superconductor mass as a function of electromagnet loop radius, *a*. The larger loop radius requires the least amount of superconducting mass. The three dots indicate three potential solutions. The annotations for each dot show estimates for the amount of bulk rock that is required to mine key elements for superconductors. For the smallest loop radius, $a \sim 10$ km, the construction project would need $\sim 10^{19}$ g of Bi, and given Solar System abundances, this amount of Bi would require mining $7 \times 10^{25}$ g of bulk rock. This amount of bulk rock is equivalent to $\sim 10\%$ of Mars. Alternatively, a larger loop radius, the radius of Mars, would require $\sim 10^{12}$ g of Bi and $\sim 5 \times 10^{18}$ g of bulk rock, or 0.1% of Olympus Mons. The most reasonable solution is a magnetic shield wrapped around Mars.

**Acknowledgement.** We would like to thank Taylor Erwin for making the beautiful graphics shown in Fig. 3 as well as the referee for their very useful comments and suggestions.

**Conflict of interest.** None.